\documentclass[sigconf, nonacm]{acmart}

\newcommand\vldbdoi{XX.XX/XXX.XX}
\newcommand\vldbpages{XXX-XXX}
\newcommand\vldbvolume{14}
\newcommand\vldbissue{1}
\newcommand\vldbyear{2022}
\newcommand\vldbauthors{\authors}
\newcommand\vldbtitle{\shorttitle} 
\newcommand\vldbavailabilityurl{URL_TO_YOUR_ARTIFACTS}
\newcommand\vldbpagestyle{plain}

\usepackage{url}
\usepackage{booktabs}
\newtheorem{proposition}{Proposition}
\newtheorem{definition}{Definition}
\newtheorem{lemma}{Lemma}
\usepackage{subfigure}
\usepackage{mathtools}
\usepackage{listings}

\usepackage{xcolor}

\definecolor{codegreen}{rgb}{0,0.6,0}
\definecolor{codegray}{rgb}{0.5,0.5,0.5}
\definecolor{codepurple}{rgb}{0.58,0,0.82}
\definecolor{backcolour}{rgb}{0.95,0.95,0.92}

\lstdefinestyle{mystyle}{
    backgroundcolor=\color{backcolour},   
    commentstyle=\color{codegreen},
    keywordstyle=\color{magenta},
    numberstyle=\tiny\color{codegray},
    stringstyle=\color{codepurple},
    basicstyle=\ttfamily\footnotesize,
    breakatwhitespace=false,         
    breaklines=true,                 
    captionpos=b,                    
    keepspaces=true,                 
    numbers=left,                    
    numbersep=5pt,                  
    showspaces=false,                
    showstringspaces=false,
    showtabs=false,                  
    tabsize=2
}

\newcommand{\figwidth}{83mm}
\newcommand{\figspace}{\vspace{-5mm}}

\begin{document}
\title{INCHE: High-Performance Encoding for Relational Databases through Incrementally Homomorphic Encryption}

\author{Dongfang Zhao}
\affiliation{%
  \institution{University of Nevada, Reno, United States}
}
\email{dzhao@unr.edu}

\begin{abstract}
Homomorphic encryption (HE) offers data confidentiality by executing queries directly on encrypted fields in the database-as-a-service (DaaS) paradigm.
While fully HE exhibits great expressiveness but prohibitive performance overhead,
a better balance between flexibility and efficiency can be achieved by partially HE schemes.
Performance-wise, however, the encryption rate of state-of-the-art HE schemes is still orders of magnitude lower than the I/O throughput,
rendering the HE scheme the performance bottleneck.

This paper proposes INCHE,
an incrementally homomorphic encryption scheme,
which aims to boost the performance of HE schemes by incrementally encrypting fields in relational databases.
The key idea of INCHE is to explore the intrinsic correlation between plaintexts and cache them for future reuse such that expensive HE primitives from plaintexts to ciphertexts are avoided.
We prove the semantic security of INCHE under the chosen-plaintext attack (CPA) model and show that its time complexity is linear in the plaintext length.
We implement an INCHE prototype by extending the Symmetria cryptosystem and verify its effectiveness on both randomly-generated data and the TPC-H benchmark.
\end{abstract}

\maketitle

\pagestyle{\vldbpagestyle}
\begingroup\small\noindent\raggedright\textbf{PVLDB Reference Format:}\\
\vldbauthors. \vldbtitle. PVLDB, \vldbvolume(\vldbissue): \vldbpages, \vldbyear.\\
\href{https://doi.org/\vldbdoi}{doi:\vldbdoi}
\endgroup
\begingroup
\renewcommand\thefootnote{}\footnote{\noindent
This work is licensed under the Creative Commons BY-NC-ND 4.0 International License. Visit \url{https://creativecommons.org/licenses/by-nc-nd/4.0/} to view a copy of this license. For any use beyond those covered by this license, obtain permission by emailing \href{mailto:info@vldb.org}{info@vldb.org}. Copyright is held by the owner/author(s). Publication rights licensed to the VLDB Endowment. \\
\raggedright Proceedings of the VLDB Endowment, Vol. \vldbvolume, No. \vldbissue\ %
ISSN 2150-8097. \\
\href{https://doi.org/\vldbdoi}{doi:\vldbdoi} \\
}\addtocounter{footnote}{-1}\endgroup

\ifdefempty{\vldbavailabilityurl}{}{
\vspace{.3cm}
\begingroup\small\noindent\raggedright\textbf{PVLDB Artifact Availability:}\\
The source code, data, and/or other artifacts have been made available at \url{https://github.com/hpdic/symmetria}.
\endgroup
}

\section{Introduction}

Database-as-a-service (DaaS)~\cite{hhaci_icde02} has now become a mainstream paradigm for data management offered by many cloud providers, such as Amazon Web Services, Google Cloud Platform, and Microsoft Azure.
Nonetheless, the concerns over outsourced data's confidentiality remains a challenging problem for applications dealing with sensitive data, e.g., patient records, financial transactions, government files.
Although various encryption schemes (e.g., AES~\cite{aes}) can be applied by the data owner before transmitting the data to the cloud provider,
a database not supporting nontrivial queries such as aggregates over the encrypted data is barely more useful than an encrypted storage server---defeating the objective of DaaS.
The common practice of adopting encryption in databases includes encrypted storage, encrypted tuples, and encrypted fields.

\textbf{Encrypted storage.} 
The database instance from the cloud vendor is considered as storage of encrypted data and the client is responsible for nontrivial queries.
This solution is viable only if (i) the relations touched by the query are sufficiently small such that the network overhead of transmitting those relations is acceptable, and (ii) the user has the capability (both computation and storage) to execute the query locally.
We stress that this solution might defeat the purpose of outsourcing the database service to the cloud.
    
\textbf{Encrypted tuples.} 
Every tuple of the original relation $R$ is encrypted into a ciphertext that is stored in column $T$ of a new relation $R^s$.
For each attribute $A_i$ in $R$, there is a corresponding attribute $A_i^s$ in $R^s$, whose value is the index of $R.A_i$.
The index is usually assigned by a random integer based on some partitioning criteria and can be retrieved with the metadata stored on the \textit{client}, i.e., the user's local node.
As a result, the schema stored at the cloud provider is $R^s(T, A_1^s, \dots, A_i^s, \dots)$.
When the user submits a query $Q$, the client splits $Q$ into two subqueries $Q_s$ and $Q_c$.
$Q_s$ serves as a filter to eliminate those unqualified tuples based on the indices in $R^s$ and transmits the qualified tuples (in ciphertexts) to the client.
$Q_c$ then ensures that those false-positive tuples are eliminated after the encrypted tuples are decrypted using the secret key presumably stored on the client.
This approach involves both the client (i.e., the user) and the server (i.e., the cloud provider) when completing a query, often referred to as \textit{information hiding} approaches~\cite{hhaci_sigmod02}.
    
\textbf{Encrypted fields.} 
The third approach aims to minimize the involvement of clients when processing the query over the encrypted data stored at the cloud provider.
The idea is encrypting the relations at a finer granularity---each attribute of a relation is separately encrypted.
The key challenge of this approach lies in its expressiveness,
e.g., how to apply arithmetic or string actions over the encrypted fields.
While fully homomorphic encryption (FHE)~\cite{cgentry_stoc09} can support a large set of computing problems,
the performance of current FHE implementations cannot meet the requirement of practical database systems~\cite{arx_vldb19,popa2011cryptdb}.
An alternative solution is partially homomorphic encryption (PHE) schemes~\cite{ppail_eurocrypt99,elgamal_tit85},
which are orders of magnitude faster than FHE but only support a single algebraic operation.
Traditional PHE schemes are designed for public/private key (asymmetric) encryption,
which is desirable for straightforward key distribution over insecure channels but significantly more expensive than secret-key (symmetric) encryption.
However, in the context of DaaS, 
the user usually serves as both the sender and the receiver and there is no need to distribute the key.
To this end, symmetric (partially) homomorphic encryption (SHE), was proposed~\cite{symmetria_vldb20,apapa_osdi16}.
This paper is along this line of research.

\textbf{Motivation.}
Although SHE delivers much faster encryption than the conventional PHE, 
both schemes assume that the underlying data is \textit{static}.
That is, they are not designed for frequently-updated data or streaming data that are commonly found in applications like video analysis~\cite{bhaynes_sigmod21,mdaum_icde21}.
If existing data are updated or new data arrive, 
the best a SHE scheme can do is to (re-)encrypt the data in its entirety.
As illustrated later in this paper~(\S\ref{sec:eval}),
the state-of-the-art SHE~\cite{symmetria_vldb20} can encrypt 32-bit random integers at a rate of 3 Mbps---much lower than the commodity network bandwidth~(cf. Fig.~\ref{fig:2022_VLDB_IncHE_Performance}) that is in the order of tens of Mbps or even Gbps.
For those types of data-intensive applications, therefore, 
it is the encryption subsystem,
rather than the I/O subsystem, 
turning to be the performance bottleneck.
This work is our first step toward boosting the performance of encrypting dynamic data in an \textit{incremental} manner.

\textbf{More related work.}
The notion of \textit{incremental cryptography} was first formalized in 1990s~\cite{mbellare_crypto94,mbellare_stoc95},
mainly from a theoretical prospective.
More recent work on incremental encryption schemes can be found in~\cite{imironov_eurocrypt12,pananth_eurocrypt17,lkhati_sac18}.
Incremental encryption recently draws a lot of research interests for efficient data encoding in the resource-constraint contexts such as mobile computing~\cite{fwang_fgcs21,gke_journal21,tbhatia_ccpe20}.
To our knowledge, however, no existing cryptosystem supports both homomorphic encryption and incremental encoding simultaneously.

\textbf{Objectives.}
Our long-term goal of this line of research is to develop a full-fledged, \underline{inc}rementally \underline{h}omomorphic \underline{e}ncryption (INCHE) scheme for streaming and frequently-updated data.
As a starting point, this paper shares our early findings as follows.

We propose an INCHE scheme in~\S\ref{sec:inche}.
Specifically, in~\S\ref{subsec:inche_overview} and \S\ref{subsec:inche_scheme}, we first sketch the intuition behind the design of INCHE after which we formalize its building blocks and protocols.
In~\S\ref{subsec:inche_security}, we prove the semantic security of INCHE if the assumption held for the adversary $\mathcal{A}$ in a conventional (batch) homomorphic encryption scheme also holds for $\widehat{\mathcal{A}}$ in INCHE.
In~\S\ref{subsec:inche_time}, we demonstrate that INCHE is theoretically efficient as its time complexity is asymptotically linear in the plaintext length.
We implement a prototype of INCHE by extending Symmetria~\cite{symmetria_vldb20}.
We evaluate INCHE with three workloads on TPC-H~\cite{tpch3} and randomly generated numbers.
Workload \#1 encrypts plaintexts in a specific field.
Workload \#2 aggregates the encrypted ciphertexts from workload \#1.
Workload \#3 encrypts plaintexts with limited memory capacity,
which can be substantiated by resource-constraint applications.
In~\S\ref{sec:eval}, we report the experimental results of comparing INCHE with Symmetria:
(i) INCHE is up to 3x faster with negligible overhead (for constructing indexes) when memory is sufficient;
(ii) INCHE is three orders of magnitude faster when aggregating ciphertexts; and
(iii) when the memory capacity is limited, INCHE is 1.3--2.1x faster.

\section{Preliminaries}
\label{sec:pre}

The notion \textit{homomorphism} originates from the study of an algebraic group,
which is an algebraic structure over a nonempty set.
Formally, a group $G$ over a set $S$ is a tuple $(G, \oplus)$,
where $\oplus$ is a binary operator satisfying the following four axioms (or properties, written in first-order logical formulae):
(i) $\forall_{g, h \in S} (g \oplus h \in S)$,
(ii) $\exists_{u \in S} \forall_{g \in S} ((g \oplus u = g) \wedge (u \oplus g = g))$;
(iii) $\forall_{g \in S} \exists_{h \in S} ((g \oplus h = u) \wedge (h \oplus g = u))$, where $u$ is defined in (ii) and $h$ is usually denoted $-g$; and
(iv) $\forall_{g, h, j \in S} ((g \oplus h) \oplus j = g \oplus (h \oplus j))$.
If we have another group $(H, \otimes)$ and a function $\varphi: G \rightarrow H$ such that 
$\forall_{g_1, g_2 \in G} (\varphi(g_1) \otimes \varphi(g_2) = \varphi(g_1 \oplus g_2))$,
then we call function $\varphi$ a homomorphism.

\textit{Homomorphic encryption} is a natural extension of group homomorphism:
Function $\varphi$ is realized as a specific encryption scheme, say $he(\cdot)$,
and the sets of plaintexts and ciphertexts correspond to the domain and the codomain of $\varphi$, respectively.
Given two plaintexts $m_1$ and $m_2$, both of which are encrypted into $c_1 = he(m_1)$ and $c_2 = he(m_2)$, respectively,
the cloud vendor can directly calculate and send $c_1 \otimes c_2$ back to the user who submits a query for $m_1 \oplus m_2$,
denoting by $\otimes$ the binary operator for ciphertexts and by $\oplus$ the binary operator for plaintexts.
The plaintext can be revealed by the user through the decrption of $he(m_1 \oplus m_2) = he(m_1) \otimes he(m_2) = c_1 \otimes c_2$.
Common operators for plaintexts include arithmetic plus $+$ and arithmetic multiplication $\times$,
although they can be defined arbitrarily.
If a homomoephic encrption scheme, say $he(\cdot)$,
supports only one binary operator in the plaintexts, 
$he(\cdot)$ is called a \textit{(partially) homomorphic encryption};
if $he(\cdot)$ supports both $+$ and $\times$ in the plaintexts,
$he(\cdot)$ is called a \textit{fully homomorphic encryption}.

Broadly speaking, encryption schemes can be categorized into two types: \textit{symmetric encryption} and \textit{asymmetric encryption}.
Symmetric encryption specifies a single secret key for both encryption and decryption;
asymmetric encryption adopts a pair of the private key and public key for encryption and decryption, respectively.
The benefit of applying asymmetric encryption is its flexibility:
The private key does not need to be distributed.
However, asymmetric encryption is orders of magnitude slower than symmetric encryption.
This work focuses on extending symmetric homomorphic encryption (e.g., Symmetria~\cite{symmetria_vldb20}) to \textit{incrementally} encode the fields of relational databases.

When developing a new encryption scheme,
it is important to demonstrate its security level, 
ideally in a provable way.
One well-accepted paradigm with a good trade-off between efficiency and security guarantee is to assume that the adversary is able to launch a \textit{chosen-plaintext attack} (CPA),
meaning that the adversary can, somehow, obtain the ciphertext of an arbitrary (i.e., chosen) plaintext.
Practically speaking, however, the adversary should only be able to obtain a polynomial number of such pairs of plaintexts and ciphertexts,
assuming the adversary's machine/algorithm takes polynomial time without unlimited computational resources.
Ideally, even if the adversary $\mathcal{A}$ can obtain those extra pieces of information, 
$\mathcal{A}$ should not make a \textit{significantly} better decision for the plaintexts than a random guess.
To quantify the degree of ``significantly better decision'',
\textit{negligible function} is introduced.
A function is called negligible if for all polynomials $poly(n)$ the following inequality $\mu(n) < \frac{1}{poly(n)}$ holds for sufficiently large $n$'s.
For completeness, we list the following lemmas for negligible functions that will be used in later sections.
We skip the proofs, which can be found in any introductory cryptography or complex theory texts.

\begin{lemma}[The summation of two negligible functions is also a negligible function]
\label{thm:neg_sum}
Let $\mu_1(n)$ and $\mu_2(n)$ be both negligible functions.
Then $\mu(n)$ is a negligible function that is defined as $\mu(n) \triangleq \mu_1(n) + \mu_2(n)$.
\end{lemma}

\begin{lemma}[The quotient of a polynomial function over an exponential function is a negligible function]\label{thm:neg_quotient}

$\frac{ploy(n)}{2^n}$ is a negligible function. 
That is,
$\exists_{N \in \mathbb{N}^*} \forall_{n \geq N} \left(\frac{ploy(n)}{2^n} < \frac{1}{poly(n)} \right)$. 
\end{lemma}

\section{INCHE: Incrementally Homomorphic Encryption}
\label{sec:inche}

\subsection{Overview}
\label{subsec:inche_overview}

The key idea of INCHE is twofold.
Firstly, we sample a polynomial number of representative values with some distribution.
In this paper, we start with a simple uniform distribution.
We precompute the encryption of those representative values before taking in a query.
Secondly, we precompute incremental deltas between two adjacent representatives such that any arbitrary value between two adjacent representatives can be represented as a summation of radix coefficients (in logarithmic time).
When encrypting a series of data of a specific attribute,
we decompose the plaintext into one representative value along with multiple deltas and leverage the homomorphic property among representatives and deltas.
Our hypothesis is that by (re)using the cached encryption of representative values and deltas,
we can avoid the relatively expensive cost of encrypting an arbitrary plaintext by directly composing the same ciphertext with the cheaper homomorphic operation.

The proposed incremental technique can be applied to arbitrary batch homomorphic encryption (HE) schemes~\cite{symmetria_vldb20,apapa_osdi16,ppail_eurocrypt99,elgamal_tit85}.
For brevity, we will assume the underlying HE scheme is \textit{symmetric}, 
i.e., the key used to encrypt the plaintext will be securely shared with the recipient who will use the same key for decryption.
Note that it is not uncommon in the context of database-as-a-service (DaaS) that a user behaves as both the sender and receiver:
Alice uses the key $K$ to encrypt the data before uploading it to a SQL Server service offered by Microsoft Azure,
and later on, Alice uses $K$ to decrypt the ciphertext hosted in the cloud.
In this case, sharing the secret key $K$ is trivial,
assuming the node storing $K$ is secure.
Nonetheless, we remark that the proposed incremental scheme can be naturally extended to \textit{asymmetric} scenarios:
We replace the secret key $K$ with the private key in asymmetric encryption.

\subsection{Scheme Description}
\label{subsec:inche_scheme}

We assume the plaintext can be encoded by $n$ bits.
For example, if we are encrypting non-negative integers,
then there can be up to $2^n$ distinct plaintexts.
We denote by $m$ the number of tuples in a relation.
Note that this implies that $m \leq 2^n$.
We denote by $poly(n)$ the set of polynomials in $n$.
If the context is clear, $poly(n)$ also refers to a specific polynomial in $n$.
We now define two important building blocks of INCHE: pivot and nuance.

\begin{definition}[Pivot]
A pivot in incremental homomorphic encryption is one plaintext whose ciphertext is precomputed and cached.
\end{definition}

\begin{definition}[Nuance]
A nuance in incremental homomorphic encryption is a pair $( \xi, he(\xi) )$,
where $\xi$ is a plaintext and $he(\xi)$ is the homomorphic encryption of $\xi$.
\end{definition}

We use $p = \Theta(poly(n))$ to denote the \textit{asymptotic} number of pivots that will be preprocessed.
Common values for $p$ include $n^c$, $1 \leq c \leq 5$~\cite{sarora_book09}.
Similarly, we use $d = \Theta(poly(n))$ to denote the \textit{asymptotic} number of nuances that will be encrypted and cached.
We denote the underlying batch HE by a 5-tuple $\Pi \triangleq (\mathcal{P}, \mathcal{C}, \mathcal{K}, \mathcal{E}, \mathcal{D})$,
where $\mathcal{P}$ is the set of plaintexts, $\mathcal{C}$ is the set of ciphertexts, $\mathcal{K}$ is the set of secret keys (since we assume the underlying encryption scheme is symmetric),
$\mathcal{E}$ and $\mathcal{D}$ are sets of keyed encryption and decryption functions and satisfy the following predicate: 
\[
\forall_{K \in \mathcal{K}} \forall_{val \in \mathcal{P}} \exists_{e_K \in \mathcal{E}} \exists_{d_K \in \mathcal{D}} (d_K(e_K(val)) = val).
\]
An incremental homomorphic cryptosystem is a 7-tuple extended from $\Pi$ denoted as
$\widehat{\Pi} \triangleq (\mathcal{P}, \mathcal{C}, \mathcal{K}, \widehat{\mathcal{E}}, \mathcal{D}, \mathcal{B}, \mathcal{N})$,
where $\mathcal{B}$ is a function from plaintexts to the set of the indexed pivots, $\mathcal{N}$ is a nuance function from a polynomial number of radix plaintexts to their ciphertexts, and $\widehat{\mathcal{E}}$ is the set of keyed functions for incremental encryption.
We detail $\mathcal{B}$, $\mathcal{N}$, and $\widehat{\mathcal{E}}$ as follows.

The value of function $\mathcal{B}(val)$ is calculated as the largest pivot that is smaller than $val$ (assuming the pivots $P_i$'s are sorted in an increasing order: $P_0 \leq P_1 \leq P_2 \leq \dots$):
$\mathcal{B}(val) \triangleq e_K(P_i)$,
where $P_i \leq val < P_{i+1}$ and $i$ is the pivot index.
The nuance function maps a logarithmic distance from $P_i$ to its encryption: 
\[
\begin{split}
\mathcal{N}: \left[1, \left\lceil \frac{P_{i+1} - P_{i}}{2} \right\rceil \right] & \rightarrow \mathcal{C}, \\
\xi & \mapsto e_K(\xi),
\end{split}
\]
where $\xi \in \left\{2^j : \forall_{j \in \mathbb{N}^*} \left(2^j \leq \left\lceil \frac{P_{i+1} - P_{i}}{2} \right\rceil \right) \right\}$.
By convention, we use $dom(\mathcal{N})$ to denote the \textit{domain} of function $\mathcal{N}$, 
i.e., the set of radix plaintexts between two adjacent pivots.
It is evident to see that 
\[
val = P_i + \sum_{j = 1}^{\vert dom(\mathcal{N}) \vert} \{0,1\} \times 2^j.
\]
We are now ready to define $\widehat{\mathcal{E}}$.
Let $e_{Ki} \in \widehat{\mathcal{E}}$, 
$\oplus$ denote the homomorphic binary summation,
and $\bigoplus$ denote the homomophic summation over a series of ciphertext summands,
then each incremental encryption function in $\widehat{\mathcal{E}}$ is calculated as follows:
\[\displaystyle
\begin{split}
e_{Ki}(val) & = e_K \left(P_i + \sum_{j = 1}^{\vert dom(\mathcal{N}) \vert} \{0,1\} \times 2^j \right) \\
    & = e_K(P_i) \oplus e_K\left(\sum_{j = 1}^{\vert dom(\mathcal{N}) \vert} \{0,1\} \times 2^j \right) \\
    & = e_K(P_i) \oplus \bigoplus_{j = 1}^{\vert dom(\mathcal{N}) \vert} e_K\left(\{0,1\} \times 2^j \right) \\
    & = \mathcal{B}(val) \oplus \bigoplus_{\xi \in dom(\mathcal{N})} \mathcal{N}(\xi) \times \{0, 1\}.
\end{split}
\]

\subsection{Semantic Security}
\label{subsec:inche_security}

This section proves the semantic security of INCHE.
Intuitively, because we only precompute and store a polynomial number $poly(n)$ of pivots and nuances (in the bit-string length $n$),
those extra pieces of information can only negligibly help the adversary---who runs a probabilistic polynomial-time (PPT) Turing machine---in the sense that the overall space is exponential $2^n$.
That is, it is computationally infeasible for a PPT adversary to break INCHE.

Technically, we want to \textit{reduce} the problem of breaking a batch homomorphic encryption scheme to the problem of breaking the incremental counterpart.
In other words, if a PPT adversary $\mathcal{A}$ takes an algorithm $alg$ to break INCHE, then $\mathcal{A}$ can efficiently (i.e., in polynomial time) construct another algorithm $alg'$ that calls $alg$ as a subroutine to break the batch encryption as well (simulating $alg'$ with $alg$).
However, if we already know that the batch encryption is semantically secure,
the above cannot happen---leading to a contradiction,
proving that INCHE would be semantically secure.
We formalize the above reasoning in the following proposition.

\begin{proposition}
If a batch homomorphic encryption $\Pi$ is semantically secure under the threat model of chosen-plaintext attack (IND-CPA),
then its corresponding extension $\widehat{\Pi}$ defined in~\S\ref{subsec:inche_scheme} is IND-CPA.
\end{proposition}

\begin{proof}

We set $\mathcal{N}(x) = \{0 \}$ in $\widehat{\Pi}$,
which implies that all the coefficients of the logarithmic distances are zero.
Because we assume a polynomial number of nuances,
this procedure takes $poly(n)$ time.
It follows that given a plaintext $val$, we have $e_{Ki}(val) = \mathcal{B}(val) = e_K(P_i)$,
where $e_{Ki}$ and $e_{K}$ are the incremental and batch encryption functions, respectively.
That is, the incremental scheme $\widehat{\Pi}$ is degraded to $\Pi$ with $P_i \in \mathcal{P}$.
This implies that $\widehat{\Pi}$ is at least as difficult as $\Pi$;
or equivalently, $\Pi$ is no harder than $\widehat{\Pi}$,
denoted by $\Pi \leq_p \widehat{\Pi}$.

Notionally, let $CPA^{\mathcal{A}}_{X}$ denote the indistinguishability experiment with scheme $X$.
The probability for $\mathcal{A}$ to successfully break $\Pi$ and $\widehat{\Pi}$ are $Pr\left[CPA_{\Pi}^\mathcal{A} = 1 \right]$ and $Pr\left[CPA_{\widehat{\Pi}}^\mathcal{A} = 1 \right]$, respectively.
By assumption, the following inequality holds:
\begin{equation}\label{eq:cpa_b}
Pr\left[CPA_{\Pi}^\mathcal{A} = 1 \right] \leq \frac{1}{2} + \epsilon,     
\end{equation}
where $\epsilon$ is a negligible probability.
By comparing $\Pi$ and $\widehat{\Pi}$,
the latter yields $p + d$ additional pairs of plaintexts and ciphertexts (out of the total $2^n$ possible pairs).
Therefore, the following inequality holds:
\begin{equation}\label{eq:cpa_diff}
    Pr\left[CPA_{\widehat{\Pi}}^\mathcal{A} = 1\right] - Pr\left[CPA_{\Pi}^\mathcal{A} = 1\right] \leq \frac{p + d}{2^n}.
\end{equation}
Combining Eq.~\eqref{eq:cpa_b} and Eq.~\eqref{eq:cpa_diff},
we have the following inequality:
\[
Pr\left[CPA_{\widehat{\Pi}}^\mathcal{A} = 1\right] \leq \frac{1}{2} + \epsilon + \frac{p + d}{2^n}
 = \frac{1}{2} + \epsilon + \frac{poly(n)}{2^n},
\]
where the last equality comes from the simple fact that the summation of two polynomials is also a polynomial:
\[
\forall_{x \in poly(n)} \forall_{y \in poly(n)} ((x + y) \in poly(n)).
\]
Now, we only need to show that the summation of the last two terms, $\epsilon + \frac{poly(n)}{2^n}$, is negligible.
According to Lemma~\ref{thm:neg_sum} and Lemma~\ref{thm:neg_quotient}~(\S\ref{sec:pre}), 
this is indeed the case.
Therefore, the probability for the adversary $\mathcal{A}$ to succeed in the $CPA_{\widehat{\Pi}}^A$ experiment is only negligibly higher than $\frac{1}{2}$,
proving the semantic security of INCHE, as claimed.
\end{proof}

\subsection{Time Complexity}
\label{subsec:inche_time}

Suppose we will encrypt a plaintext $val$ from a specific field and the index pivots are managed in a B+ tree.
It takes $O(\log p)$ to locate an appropriate pivot $P_i$ at a leaf node such that $P_i \leq val$ and $P_{i+1} > val$.
Let $\Delta P$ denote the range between $P_i$ and $P_{i+1}$, $0 \le i < p$,
assuming the $P_i$'s are selected with equal widths.
It follows that there exist $\left\lceil \log_2 \Delta P \right\rceil$ nuances,
denoted $\xi_j = \left(\xi_j^p, \xi_j^c \right) = \left( \left\lceil \frac{\Delta P}{2^j} \right\rceil, he\left( \left\lceil \frac{\Delta P}{2^j} \right\rceil \right) \right)$,
$1 \leq j \leq \left\lceil \log_2 \Delta P \right\rceil$.
Because there exist up to $2^n$ distinct numbers given a bit-string of length $n$, 
it holds that $\Delta P \leq \frac{2^n}{p}$.
It follows that 
\[
\left\lceil \log_2 \Delta P \right\rceil \leq \log_2 \Delta P + 1 \leq \log_2 \frac{2^n}{p} + 1
    = n- \log_2 p + 1.
\]
Consequently, the time complexity of encrypting a single plaintext with INCHE is asymptotically linear in the length of its bit-string:
\[
\log p + \left\lceil \log \Delta P \right\rceil \leq \log p + n - \log p + 1 = O(n).
\]

\section{Experimental Results}
\label{sec:eval}

\textbf{Implementation.}
We implement INCHE with Java by extending Symmetria~\cite{symmetria_vldb20},
which serves as the baseline in our experiments.\footnote{Seabed~\cite{apapa_osdi16} is another symmetric homomorphic cryptosystem but only supports homomorphic addition (e.g., no subtraction or negation) and is not open-source.}
The pivots are implemented as a hash tree:
The pivots are keys with the encrypted values as the data records.
The nuances are implemented with a hash table,
whose keys represent the plaintext deltas and values are the homomorphic encryption of the keys.
The source code is currently hosted on Github.com as a branch of the Symmetria codebase~\cite{symmetria_github}: \url{https://github.com/hpdic/symmetria}.
We will construct a website for further development and updates. 

\textbf{Testbeds.}
The prototype is evaluated on two testbeds.
Most experiments are carried out on a Lenovo workstation with Intel(R) Core(TM) i7-6820HQ CPUs, 64~GB DDR4 RAM, and one Samsung PCIe NVME SSD of 1~TB.
Unless otherwise stated, results are collected from the Lenovo workstation.
Expensive experiments (e.g., ciphertext aggregation) are conducted on CloudLab~\cite{cloudlab}.
We use the \texttt{c6420} instances,
each of which is equipped with two 16-core Intel Xeon Gold 6142 CPUs at 2.6 GHz, 384 GB ECC DDR4-2666 memory, and 	
two Seagate 1~TB 7200 RPM 6G SATA HDDs.
The operating system image is Ubuntu 20.04.3 LTS.

\textbf{Data sets.}
Two data sets will be used in the following evaluation.
The first data set is the TPC-H benchmark, version 3.0.0~\cite{tpch3}.
We generate the tables with various scales up to 100 (i.e., ``$-s\;100$''),
constituting roughly a 100 GB relational database.
The second data set is a set of numbers randomly generated from $\left[0, 2^{64}\right)$.
The benefit of working with the second data set is that we can arbitrarily control the parameters of the data, such as the length of the bit-string and distribution of the values.
For all experiments, we repeat the executions at least three times and report the average and the standard deviation (\texttt{stdev} or \texttt{error}).

\subsection{Encoding TPC-H Relations}

\begin{figure}[!t]
    \centering
    \includegraphics[width=\figwidth]{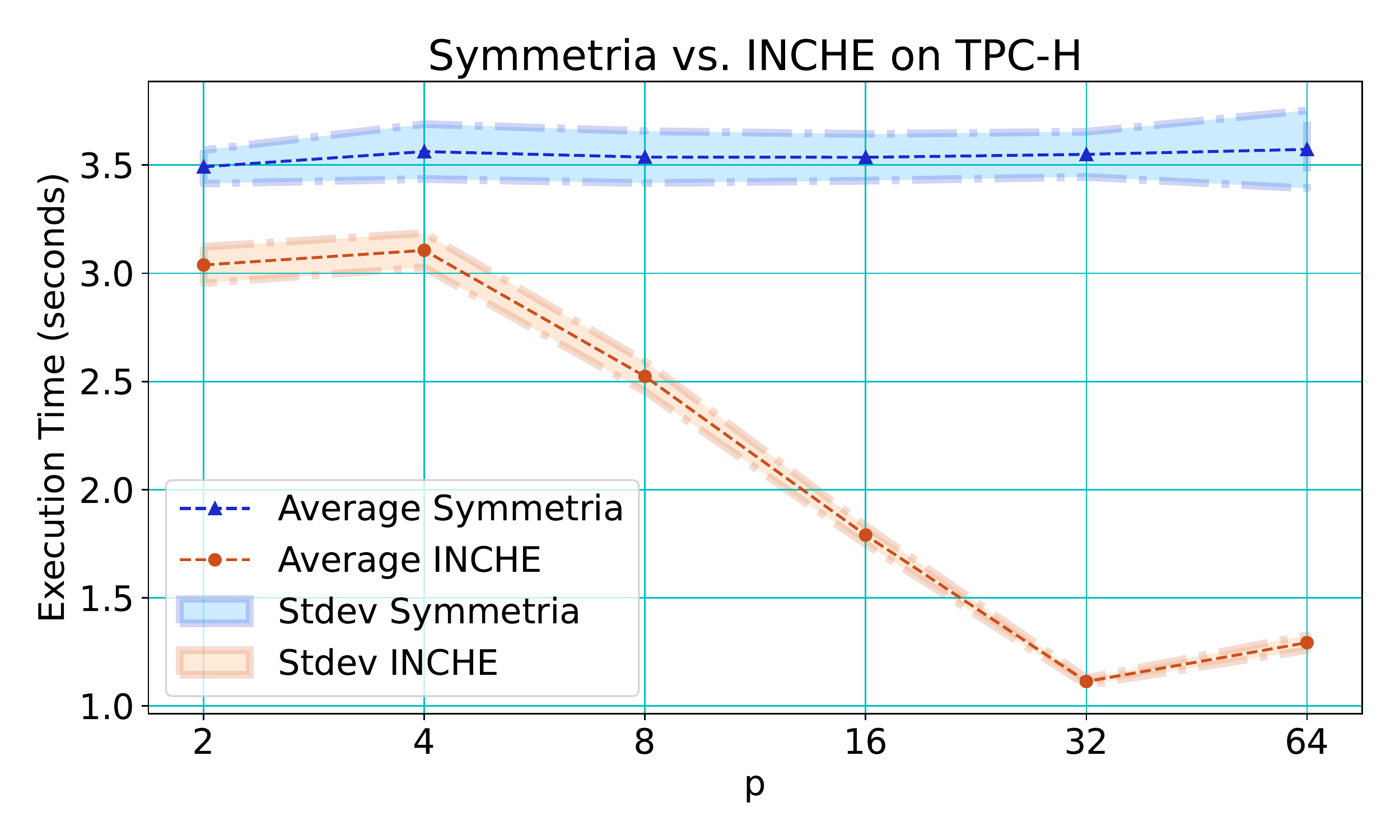}
    \caption{Performance comparison on TPC-H, scale = 100, 20,000,000 tuples in table Part.}
    \label{fig:2022_VLDB_IncHE_TPCH_Comp}
     \figspace
\end{figure}

We start by applying INCHE to the Part.P\_Size attribute.
With the option ``-s 100'', there are overall 20,000,000 tuples in the Part table.
We vary the number of pivots (i.e., $p$) in the $x$-axis between 2 and 64.
We report the performance of INCHE (without the overhead of constructing the $p$-tree and $d$-hash, which will be reported in the next experiment), and compare it against Symmetria in Fig.~\ref{fig:2022_VLDB_IncHE_TPCH_Comp}.
Generally speaking, larger $p$ values allow INCHE to complete faster because of the finer granularity of the gaps among $p$'s as well as fewer nuances.
Notably, INCHE is about 3x faster than Symmetria when $p = 32$.
If the plaintexts are \textit{overly} split (e.g., $p = 64$), the extra cost for maintaining the $p$-tree may outweigh the benefit of $d$-hash,
causing performance suboptimal.

\begin{figure}[!t]
    \centering
    \includegraphics[width=\figwidth]{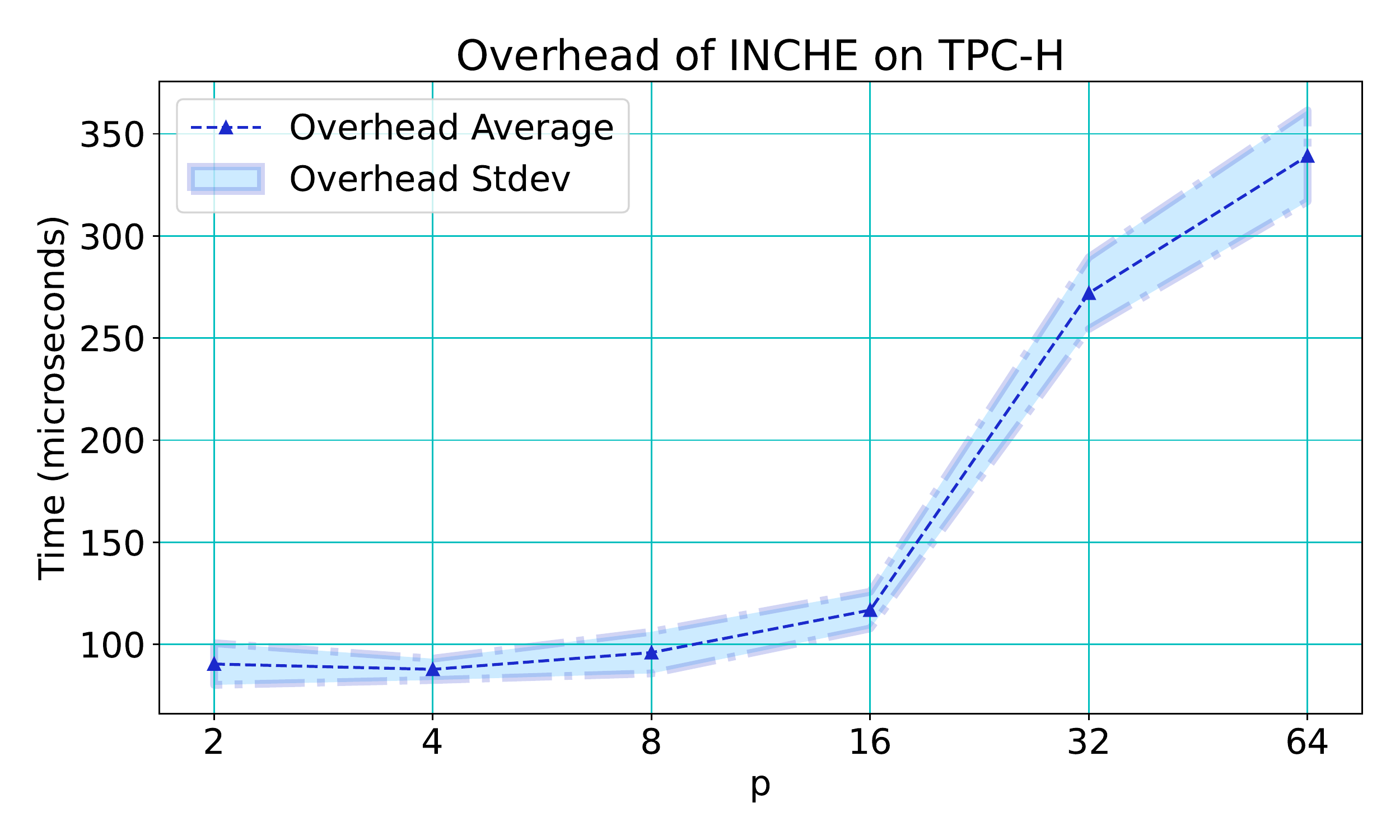}
    \caption{INCHE overhead on TPC-H, scale = 100, 20,000,000 tuples in table Part.}
    \label{fig:2022_VLDB_IncHE_TPCH_Overhead}
     \figspace
\end{figure}

We report the overhead of INCHE in Fig.~\ref{fig:2022_VLDB_IncHE_TPCH_Overhead}.
We do not show the overhead in the previous experiment because the overhead is orders of magnitude smaller than the encryption of both Symmetria and INCHE:
The time for precomputing the $p$-tree and $d$-hash is in the order of sub-millisecond, from less than 100 microseconds to about 350 microseconds for $p \in [2, 64]$.
It should be noted that the overhead itself increases proportionally to the choice of $p$ due to the additional computation (and caching) of pivots.

\subsection{Encoding Randomly Generated Numbers}

\begin{figure}[!t]
    \centering
    \includegraphics[width=\figwidth]{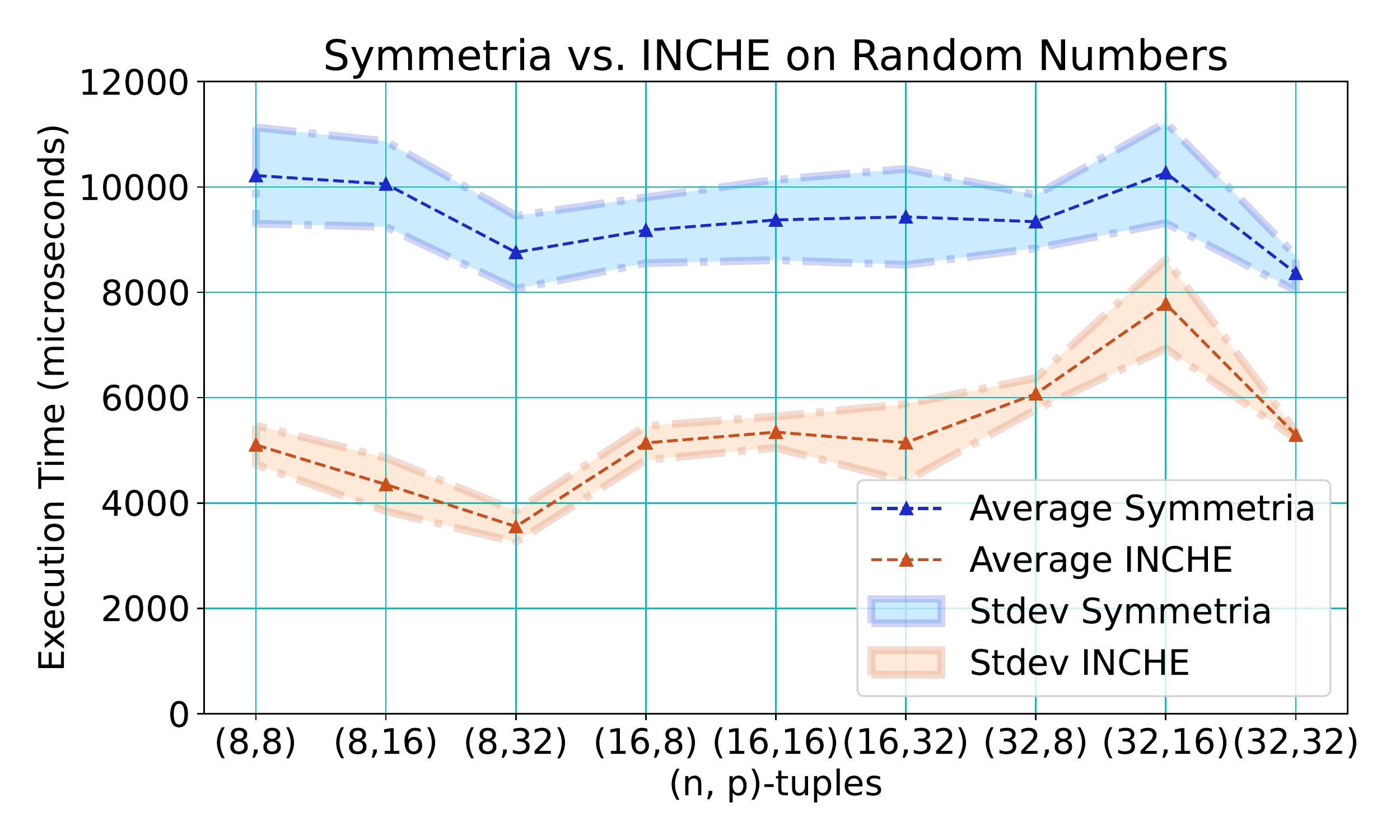}
    \caption{Performance Comparison of Symmetria and INCHE on 1,024 random plaintexts.}
    \label{fig:2022_VLDB_IncHE_Performance}
     \figspace
\end{figure}

We compare the performance of Symmetria and INCHE when encrypting 1,024 random numbers of variable lengths in Fig.~\ref{fig:2022_VLDB_IncHE_Performance}.
We in the $x$-axis vary the $(n,p)$ pairs ranging between 8 and 32,
where $n$ indicates the bit-string length and $p$ indicates the number of pivots, respectively.
We observe that INCHE consistently outperforms Symmetria for all $(n,p)$ pairs by up to 50\%,
which is aligned with the observation from TPC-H in Fig.~\ref{fig:2022_VLDB_IncHE_TPCH_Comp}.

\begin{figure}[!t]
    \centering
    \includegraphics[width=\figwidth]{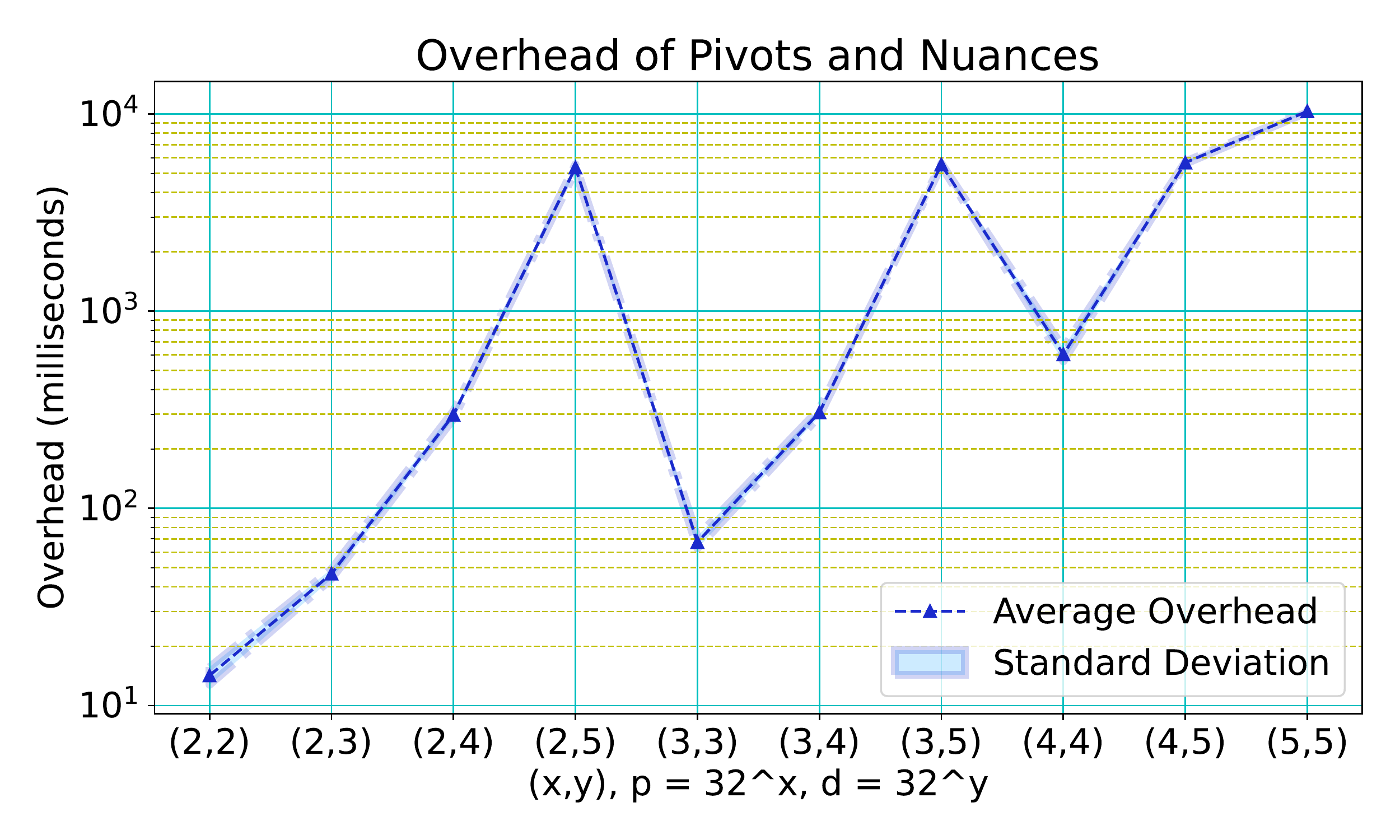}
    \caption{Performance overhead incurred by pivots and nuances when encrypting $2^{32}$ random plaintexts.}
    \label{fig:2022_VLDB_IncHE_Overhead}
     \figspace
\end{figure}

We measure the time overhead for precomputing pivots and nuances of $2^{32}$ random values.
Note that this experiment has a much larger data set than that in Fig.~\ref{fig:2022_VLDB_IncHE_Performance} (i.e., 1,024 = $2^{10}$),
because we will to a large extent vary both the number of pivots $p = n^x$, $2 \leq x \leq 5$ ($x$ is considered as a practical upper bound in complexity theory~\cite{sarora_book09}), and the number of nuances $d = n^y$, $x \leq y$.
We set $n = 32$, meaning that there are potential $2^{32}$ distinct values in the underlying data set.
The $x$-axis of Fig.~\ref{fig:2022_VLDB_IncHE_Overhead} enumerates those $(x,y)$ pairs.
We observe that although the plot shows a somewhat zig-zag pattern from small to large pairs,
the segments of fixed $d$'s are consistent with the TPC-H results (cf. Fig.~\ref{fig:2022_VLDB_IncHE_TPCH_Overhead}).

\subsection{Aggregating Encrypted Fields}

The results in this section are collected from CloudLab.
For a simple aggregate query shown in Listing~\ref{lst:sql_agg},
the execution on the scale-10 TPC-H relations computes according to the following equation:
\[
e_k\left(\sum_{i=1}^{2,000,000} s_i \right) = \bigoplus_{i=1}^{2,000,000} e_K(s_i),
\]
where $s_i$ denotes the value of the P\_Size field of the $i$-th row of relation Part.
Directly adding up $e_K(s_i)$ is costly because $\oplus$ on ciphertexts is an expensive number-theoretical operation.
INCHE allows us to cache the ciphertexts of both pivot and nuance along with their frequencies in plaintexts.
Therefore, we can reduce the frequency of $\oplus$ by $\times$ if the HE scheme supports it (Symmetria~\cite{symmetria_vldb20} does) and calculate the result as the following equation: 
\[
e_k\left(\sum_{i=1}^{2,000,000} s_i \right) = 
freq^p_{i} \times \bigoplus_{i=1}^p e_K(P_i) + freq^{\xi}_j \times \bigoplus_{j=1}^d \xi_j,
\]
where $p$ and $d$ are much smaller than 200,000 (e.g., $p = d = 32$), $freq^y_x$ indicates the frequency of the $x$-th element in the $y$-container,
and $e_K(P_i)$'s are part of the $p$-tree entries cached in memory.

\begin{lstlisting}[language={SQL}, label={lst:sql_agg}, caption={A simple SQL aggregate query on TPC-H.}]
--  TPC-H 3.0.0, "dbgen -s 10"
SELECT  AVG(P_Size)
FROM    Part;
\end{lstlisting}

\begin{figure}[!t]
    \centering
    \includegraphics[width=\figwidth]{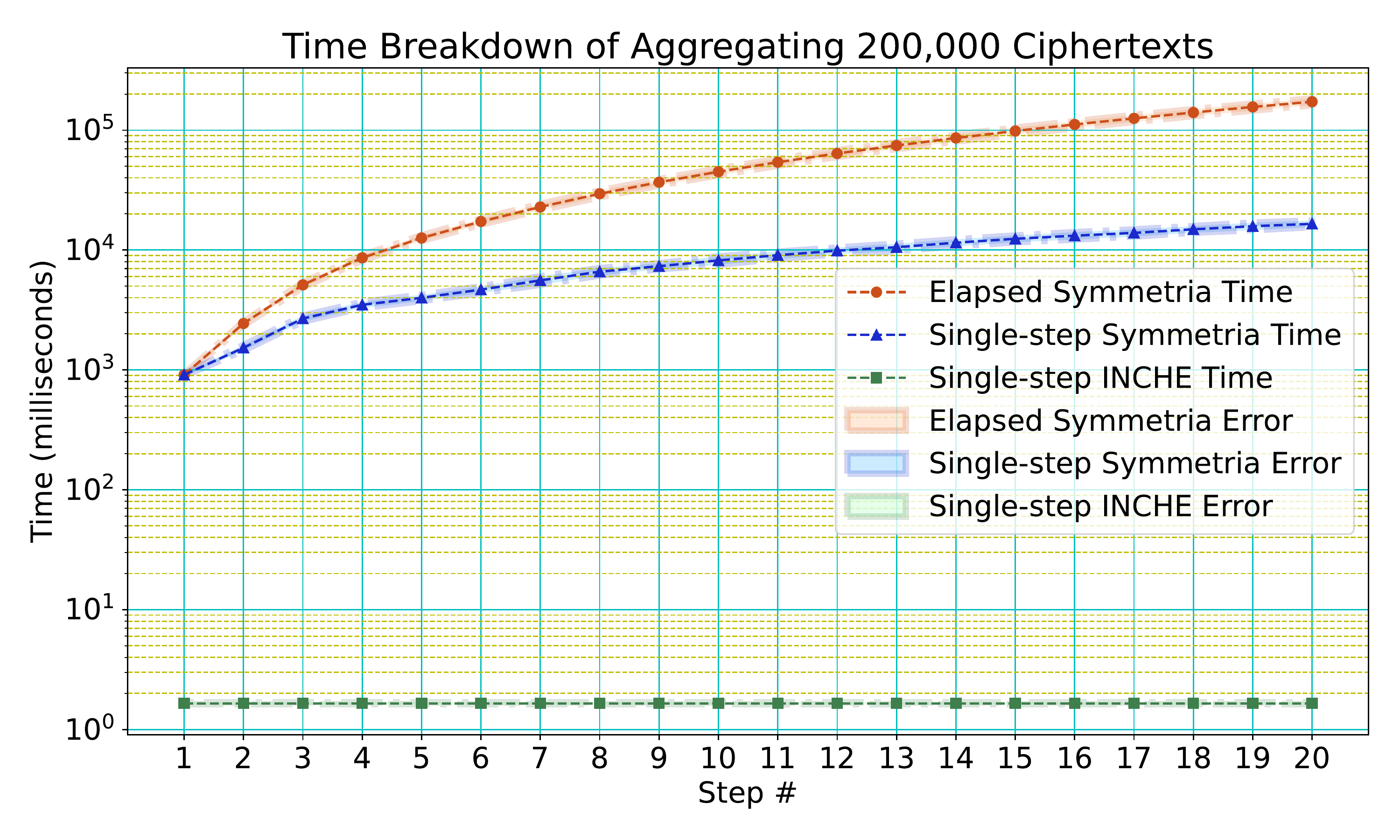}
    \caption{Time breakdown of aggregating 200,000 tuples of table Part in TPC-H.
    }
    \label{fig:2022_VLDB_IncHE_Grow}
     \figspace
\end{figure}

\begin{figure}[!t]
    \centering
    \includegraphics[width=\figwidth]{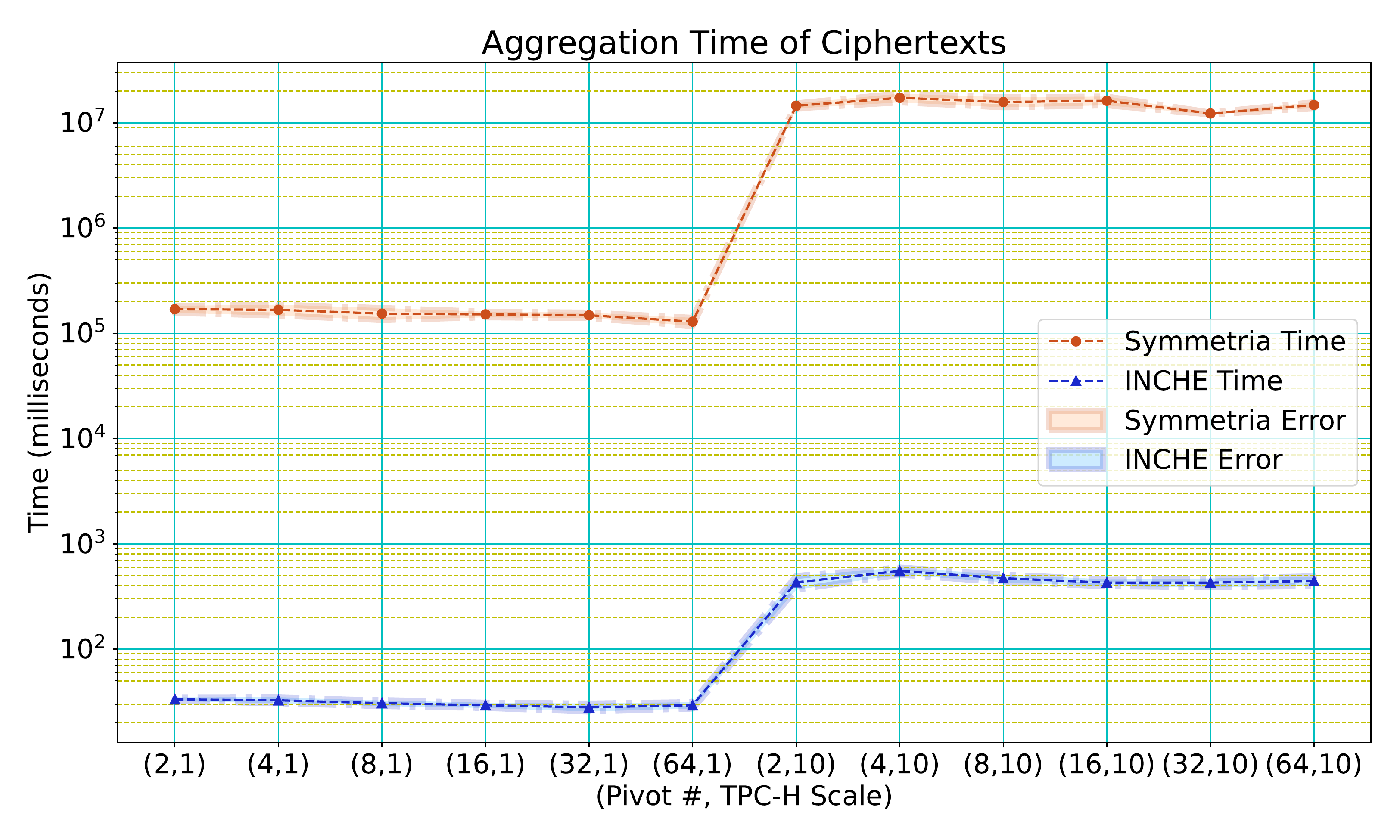}
    \caption{Aggregating time with different numbers of pivots on different TPC-H scales.
    }
    \label{fig:2022_VLDB_IncHE_Aggr}
     \figspace
\end{figure}

Fig.~\ref{fig:2022_VLDB_IncHE_Grow} reports the time for aggregating 200,000 Part.P\_Size fields in scale-1 TPC-H,
where one step comprises 10,000 encrypted fields.
We observe that the one-step cost of Symmetria is not constant:
At a larger step, it takes a longer time to aggregate the same number of new ciphertexts.
This is concerning because it implies that the batch HE scheme is not scalable.
To investigate how bad it could become,
Fig.~\ref{fig:2022_VLDB_IncHE_Aggr} reports the same workload on TPC-H of scales-1 and scale-10;
we did not report the scale-100 results because Symmetria finished only 53\% (i.e., 10,550,000 out of 20,000,000) ciphertext additions after 100 hours of execution. 
We observe that INCHE can aggregate 2,000,000 fields within a second while Symmetria takes hours to complete the same workload.

\subsection{Computing Nuances On-the-Fly}

\begin{figure}[!t]
    \centering
    \includegraphics[width=\figwidth]{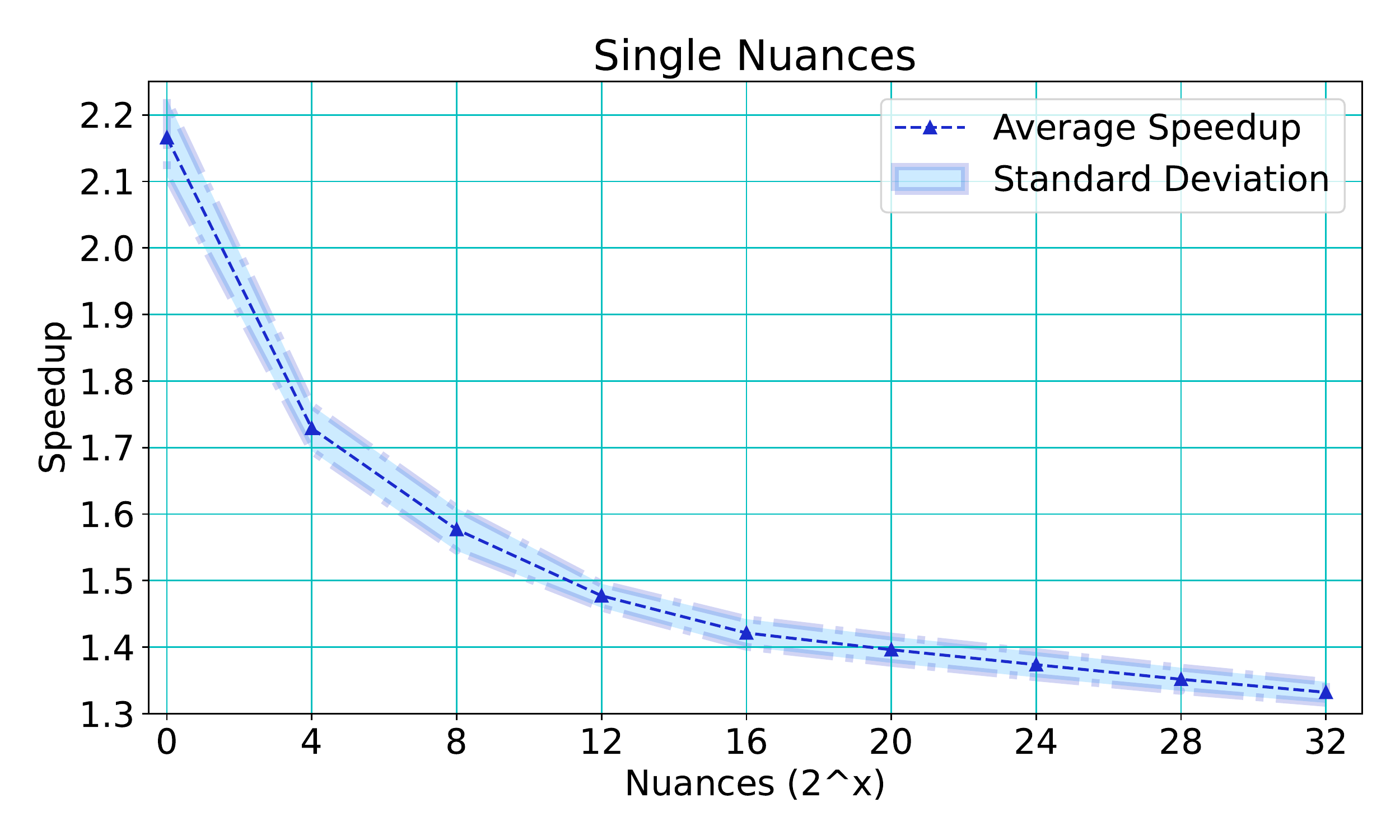}
    \caption{INCHE Speedup over Symmetria when computing nuances on-the-fly.}
    \label{fig:2022_VLDB_IncHE_SingleNuance}
     \figspace
\end{figure}

The previous sections assume that there is sufficient memory capacity to accommodate $p$ pivots and $d$ nuances.
In certain application scenarios (e.g., edge computing~\cite{aalmamun_ndss20}, supply chains~\cite{hshen_ndss20}, system-on-chip~\cite{scharles_isvlsi20}),
we might have limited resources and may not be able to hold, say, $2^{32}$ nuances as in Fig.~\ref{fig:2022_VLDB_IncHE_TPCH_Overhead}.
Therefore, the following experiment will investigate the worst-case scenario where we are forced to compute nuances on the fly.
We report the performance of adopting a single nuance for a random value in $\left[0, 2^{64}\right)$ in Fig.~\ref{fig:2022_VLDB_IncHE_SingleNuance}.
The worst-case overhead of calculating a single nuance leads to as low as 1.3x speedup over the vanilla Symmetria encryption.
In the best case, i.e., when nuance is set to one,
the speedup is over 2.1x.

\section{Conclusion and Future Work}

This paper presents INCHE, an incremental extension of an arbitrarily (batch) homomorphic encryption scheme.
Theoretically, INCHE is proven semantically secure with high efficiency, i.e., linear time complexity in the length of input bit-string.
When experimentally evaluated on TPC-H and randomly-generated plaintexts, INCHE significantly outperforms conventional homomorphic encryption:
1.3--3x faster for encryption and three orders of magnitude faster for aggregation.
Given those promising preliminary results, we plan to evaluate INCHE more extensively with real-world applications/data used in our recent database work~\cite{aalmamun_icde21,parmita_vldb17} and explore the following directions along this line of research.

\textbf{Confidential blockchains.}
While a blockchain (e.g., Bitcoin~\cite{bitcoin}, Ethereum~\cite{ethereum}) by itself exhibits strong data integrity/authenticity through (expensive) consensus protocols among the participating nodes,
the transaction data are stored in plaintexts.
That is, although the parties involved in a blockchain transaction are anonymized through (hashed) public keys,
the data (e.g., funds to be transferred) in a transaction are \textit{not} encrypted,
which are vulnerable to malicious side-channel attacks or illicit/inappropriate activities regarding privacy.
If INCHE can be adopted by blockchains,
we would be able to achieve both the \textit{confidentiality} and the \textit{integrity} of blockchain transactions.
We will first integrate INCHE to Blocklite~\cite{xwang_cloud19} for emulation on the public cloud and then to BAASH~\cite{aalmamun_sc21} for scientific computing and applications.

\textbf{Relational-algebraic extension.}
We plan to extend relational algebra (RA) with a set of INCHE primitives.
Note that some RA operators can be naturally extended to handle INCHE relations, e.g., $\cup, \cap, -$.
As a concrete example, let $R^*$ and $S^*$ be two INCHE relations (i.e., whose fields are encrypted with an INCHE scheme), it is evident that $t \in (R \cap S) \iff e_K(t) \in (R^* \cap S^*)$.
If one relation is in plaintext, we can apply the INCHE scheme before the operation.
For unitary RA operators like $\sigma$, $\pi$, and $\rho$, INCHE can be extended by encrypting the arguments (e.g., searchable encryption~\cite{dsong_sp00}).
For example, a query $\sigma_{a = 'NSF'}(R)$ can be extended into $\sigma^*_{a = e_K('NSF')}(R^*)$.
Extending INCHE schemes to joins, however, is more challenging~\cite{fhahn_icde19} unless the join is implemented by a naive composition of $\bowtie$'s and $\sigma$,
which will likely incur huge overhead and needs further research.

\bibliographystyle{abbrv}
\bibliography{ref_new}

\end{document}